\documentclass[preprint,12pt,fleqn]{elsarticle}
\usepackage{amssymb}
\usepackage{amsfonts}
\usepackage[english]{babel}
\usepackage{euscript}
\usepackage{bbm}
\usepackage{xfrac}
\usepackage{color}

\newcommand{\fl}{\hspace*{-\mathindent}}
\newcommand{\textfrac}[2]{\textstyle{\frac{#1}{#2}}}
\newcommand{\eqref}[1]{(\ref{#1})}
\newcommand{\qq}{\mathfrak{q}}
\newcommand{\ww}{\mathfrak{w}}

\begin{document}

\begin{frontmatter}

\title{
Isospectral deformation of the reduced quasi-classical self-dual Yang--Mills equation
}

\author{Oleg I. Morozov}
\ead{morozov{\symbol{64}}agh.edu.pl}
\address{Faculty of Applied Mathematics,AGH University of Science and Technology,
         \\
        Mickiewicza 30, Cracow 30-059, Poland}

\begin{abstract}
We derive new four-dimensional partial differential equation with the isospectral Lax representation
by shrinking the symmetry algebra of the reduced qua\-si-classical self-dual Yang--Mills equation.
Then we find a recursion operator for the obtained equation and construct B{\"a}cklund transformations
between this equation and the reduced qua\-si-classical self-dual Yang--Mills equation
as well as the four-dimensional Mart{\'{\i}}nez Alonso--Shabat equation
\end{abstract}


\end{frontmatter}

\section{Introduction}

Integrable partial differential equations ({\sc pde}s) play an important role in modern physics and mathematics.
While there are many non-equivalent de\-fi\-ni\-ti\-ons of integrability (see discussion in
\cite{Zakharov1991,Mikhailov2009}), the most universal perspective seems to be one based on the construction
of Lax representations, which is the starting point for
a number of powerful techniques for studying {\sc pde}s such as inverse scattering trans\-for\-ma\-ti\-ons for solitonic
equations, B{\"a}cklund transformations, recursion operators, nonlocal symmetries and conservation laws, Darboux
transformations, see \cite{WE,Zakharov82,NovikovManakovPitaevskiyZakharov1984,%
Konopelchenko1987,AblowitzClarkson1991,MatveevSalle1991,Olver1993,BacklundDarboux2001}
and references therein.
The systematic and convenient framework for dealing with these nonlocal geometric
structures is pro\-vi\-ded by the theory of differential coverings introduced by A.M. Vinogradov
in \cite{Vinogradov1982} and then elaborated in \cite{KrasilshchikVinogradov1984,KrasilshchikVinogradov1989},
see also \cite{VKL1986,VK1999,KrasilshchikVerbovetskyVitolo2017}.

From both theoretical and practical viewpoints, of the most significance are {\sc pde}s that admit isospectral Lax
representations, in other words, differential coverings with non-removable  (spectral) parameter.
Such equations are quite rare, espe\-ci\-al\-ly in the multi-dimensional case, i.e., when the number of independent
variables is greater than 2. For examples of four-dimensional {\sc pde}s with iso\-spect\-ral Lax representations
see \cite{FerapontovKhusnutdinova2004,DoubrovFerapontov2010,FerapontovKhusnutdinovaKlein2011,IgoninMarvan2014,%
Morozov2014,MorozovSergyeyev2014,CalderbankKruglikov2016,PavlovStoilov2017,DoubrovFerapontovKruglikovNovikov2017}
and references therein.

The challenging unsolved problem in the theory of integrable equations is to find conditions that are formulated
in inherent terms of a {\sc pde} under study and ensure existence of a Lax representation. Recently, an approach
to this problem has been proposed in \cite{Morozov2017,Morozov2018a,Morozov2019}, where we show that for
some {\sc pde}s their Lax representations can be inferred from the second twisted cohomology group of the contact
symmetry algebras. In paricular, paper \cite{Morozov2019} contains examples exhibiting that for some multi-dimensional
{\sc pde}s their isospectral Lax representations are related to the distinguished structure of the symmetry algebras.
Namely, in these examples the symmetry algebra of the {\sc pde} is of the form
$\mathfrak{a}_{\diamond} \ltimes (\mathbb{R}_n[h]\otimes \mathfrak{a}_\infty)$,
where $\mathfrak{a}_{\diamond}$ is a finite-dimensional Lie algebra with nontrivial second twisted cohomology
group, $\mathfrak{a}_{\infty}$ is an infinite-dimensional Lie algebra, and
$\mathbb{R}_n[h] = \mathbb{R}[h]/(h^{n+1}=0)$ is the (commutative associative unital) algebra of truncated
polynomials  of the formal variable $h$. Moreover, in \cite{Morozov2019} we show that the series of extensions
of the symmetry algebra generated by maps
$\mathbb{R}_{n}[h] \mapsto \mathbb{R}_{n+1}[h] \mapsto \mathbb{R}_{n+2}[h] \mapsto \dots$
produces the integrable hierarchy associated with the {\sc pde} under the study. It is natural to ask what
happens when one shrinks the symmetry algebra of this type via replacing $\mathbb{R}_n[h]$ by
$\mathbb{R}_{n-1}[h]$.
This idea was exploited in  \cite{KruglikovMorozov2015}, where for some four-dimensional integrable
Monge--Amp{\`{e}}re equations of Hirota type we have derived a number of their ``symmetric deformations'',
that is, equations whose symmetry algebras are constructed by omitting some graded components of
the symmetry algebra of the initial equation. The obtained equations turn out to admit Lax representations of
both isospectral and and non-isospectral types.

In the present paper we combine the techniques of \cite{KruglikovMorozov2015} and
\cite{Morozov2017,Morozov2018a,Morozov2019} and  consider the reduced quasi-classical self-dual Yang--Mills
equation
\begin{equation}
u_{yz} = u_{tx}+u_y\,u_{xx}-u_x\,u_{xy},
\label{FKh4}
\end{equation}
which admits the isospectral Lax representation
\begin{equation}
\left\{
\begin{array}{lcl}
s_t &=& \lambda\,s_y-u_y\,s_x,
\\
s_z &=& (\lambda-u_x)\,s_x
\end{array}
\right.
\label{FKh4_covering_lambda}
\end{equation}
derived in \cite{FerapontovKhusnutdinova2004}.
In \cite{Morozov2019} we show that this Lax representation
can be inferred from the structure of the Lie algebra of contact symmetries of this equation. This algebra is
isomorphic to the semi-direct product $\qq_3 = \qq_{\diamond} \ltimes \qq_{3,\infty}$ of the 6-dimensional
Lie algebra $\qq_\diamond$ and the infinite-dimensional ideal $\qq_{3,\infty}$, which is the tensor product
$\mathbb{R}_2[h] \otimes \ww[t,z]$ of the (associative commutative unital) algebra of truncated polynomials
$\mathbb{R}_2[h]= \mathbb{R}[h]/(h^3=0)$ and the Lie algebra
$\ww[t,z] = \langle t^i z^j \,\partial_z \,\,\vert \,\, i, j \in \mathbb{N}_0 \rangle$.
The subalgebra $\qq_\diamond$ has one-dimensional second twisted cohomology group, and the non-trivial 2-cocycle
generates an extension $\widehat{\qq}_3$ for $\qq_3$. The Maurer--Cartan ({\sc mc}) forms of
this extension provide the Wahlquist--Estabrook form for the Lax representation \eqref{FKh4_covering_lambda}.

In this paper we construct  the {\sc pde} that is defined by
the shrunk Lie algebra $\qq_2 =\qq_\diamond \ltimes \left(\mathbb{R}_2[h] \otimes \ww[t,z]\right)$.
This Lie algebra  admits the  extension $\widehat{\qq}_2$ generated by non-trivial twisted 2-cocycle of
$\qq_\diamond$. We show that the {\sc mc} forms of $\widehat{\qq}_2$ provide Lax representations that define
equation
\begin{equation}
u_{ty}  = u_y \,u_{xz} - u_z\,u_{xy} +y\,u_{yz}.
\label{deformed_MASh4}
\end{equation}
This equation, to the best of our knowledge, has not yet appeared in the li\-te\-ra\-tu\-re.
We show that one of the the Lax representations of \eqref{deformed_MASh4} contains non-removable spectral
parameter. This allowed us to find a recursion ope\-ra\-tor for symmetries of \eqref{deformed_MASh4}.

By the construction, equation \eqref{deformed_MASh4} can be considered as an integrable deformation of
equation \eqref{FKh4}. Notice that \eqref{deformed_MASh4} is not invariant with respect to the translation
$y \mapsto y+\epsilon$, while \eqref{FKh4} admits this transformation as a symmetry.  Likewise, equation
\eqref{deformed_MASh4} can be considered as a deformation of the four-dimensional Mart{\'{\i}}nez
Alonso--Shabat equation \cite{MartinezAlonsoShabat2002,MartinezAlonsoShabat2004,Morozov2014}
\begin{equation}
u_{ty}  = u_y \,u_{xz} - u_z\,u_{xy}.
\label{MASh4}
\end{equation}
This equation is related to equation \eqref{FKh4} by the B\"acklund transformation \cite{KruglikovMorozov2015}.
We construct a B\"acklund transformation between equations \eqref{deformed_MASh4} and \eqref{FKh4} and thus we
show that equations \eqref{deformed_MASh4} and \eqref{MASh4} are related by a B\"acklund transformation.
Notice that the contact symmetry algebras of equations \eqref{FKh4}, \eqref{deformed_MASh4}, and \eqref{MASh4}
are pairwise non-isomorphic, therefore these equations are pairwise non-equivalent with respect to the
pseudogroup of contact transformations.

\section{Preliminaries}
\label{Preliminaries_section}

\subsection{Symmetries and differential coverings}
\label{symm_dc_subsection}

The presentation in this subsection closely follows
\cite{KrasilshchikVerbovetsky2011,KrasilshchikVerbovetskyVitolo2012},
see also \cite{KrasilshchikVinogradov1984,KrasilshchikVinogradov1989,VK1999}.
Let $\pi \colon \mathbb{R}^n \times \mathbb{R}^m \rightarrow \mathbb{R}^n$,
$\pi \colon (x^1, \dots, x^n, u^1, \dots, u^m) \mapsto (x^1, \dots, x^n)$, be a trivial bundle, and
$J^\infty(\pi)$ be the bundle of its jets of the infinite order. The local coordinates on $J^\infty(\pi)$ are
$(x^i,u^\alpha,u^\alpha_I)$, where $I=(i_1, \dots, i_n)$ are multi-indices, and for every local section
$f \colon \mathbb{R}^n \rightarrow \mathbb{R}^n \times \mathbb{R}^m$ of $\pi$ the corresponding infinite jet
$j_\infty(f)$ is a section $j_\infty(f) \colon \mathbb{R}^n \rightarrow J^\infty(\pi)$ such that
$u^\alpha_I(j_\infty(f))=\displaystyle{\frac{\partial ^{\#I} f^\alpha}{\partial x^I}}
=\displaystyle{\frac{\partial ^{i_1+\dots+i_n} f^\alpha}{(\partial x^1)^{i_1}\dots (\partial x^n)^{i_n}}}$.
We put $u^\alpha = u^\alpha_{(0,\dots,0)}$. Also, we will simplify notation in the following way, e.g., in the
case of $n=4$, $m=1$: we denote $x^1 = t$, $x^2= x$, $x^3= y$, $x^4=z$ and
$u^1_{(i,j,k,l)}=u_{{t \dots t}{x \dots x}{y \dots y}{z \dots z}}$ with $i$  times $t$,
$j$  times $x$, $k$  times $y$, and $l$ times $z$.

The  vector fields
\[
D_{x^k} = \frac{\partial}{\partial x^k} + \sum \limits_{\# I \ge 0} \sum \limits_{\alpha = 1}^m
u^\alpha_{I+1_{k}}\,\frac{\partial}{\partial u^\alpha_I},
\qquad k \in \{1,\dots,n\},
\]
$(i_1,\dots, i_k,\dots, i_n)+1_k = (i_1,\dots, i_k+1,\dots, i_n)$,  are called {\it total derivatives}.
They com\-mu\-te everywhere on $J^\infty(\pi)$.

The {\it evolutionary vector field} associated to an arbitrary vector-valued smooth function
$\varphi \colon J^\infty(\pi) \rightarrow \mathbb{R}^m $ is the vector field
\[
\mathbf{E}_{\varphi} = \sum \limits_{\# I \ge 0} \sum \limits_{\alpha = 1}^m
D_I(\varphi^\alpha)\,\frac{\partial}{\partial u^\alpha_I}
\]
with $D_I=D_{(i_1,\dots\,i_n)} =D^{i_1}_{x^1} \circ \dots \circ D^{i_n}_{x^n}$.

A system of {\sc pde}s $F_r(x^i,u^\alpha_I) = 0$ of the order $s \ge 1$ with $\# I \le s$, $r \in \{1,\dots, R\}$
for some $R \ge 1$, defines the submanifold
$\EuScript{E}=\{(x^i,u^\alpha_I)\in J^\infty(\pi)\,\,\vert\,\,D_K(F_r(x^i,u^\alpha_I))=0,\,\,\# K\ge 0\}$
in $J^\infty(\pi)$.

A function $\varphi \colon J^\infty(\pi) \rightarrow \mathbb{R}^m$ is called a {\it (generator of an
infinitesimal) symmetry} of equation $\EuScript{E}$ when $\mathbf{E}_{\varphi}(F) = 0$ on $\EuScript{E}$. The
symmetry $\varphi$ is a solution to the {\it defining system}
\begin{equation}
\ell_{\EuScript{E}}(\varphi) = 0,
\label{defining_eqns}
\end{equation}
where $\ell_{\EuScript{E}} = \ell_F \vert_{\EuScript{E}}$ with the matrix differential operator
\[
\ell_F = \left(\sum \limits_{\# I \ge 0}\frac{\partial F_r}{\partial u^\alpha_I}\,D_I\right).
\]
The {\it symmetry algebra} $\mathrm{Sym} (\EuScript{E})$ of equation $\EuScript{E}$ is the linear space of
solutions to  (\ref{defining_eqns}) endowed with the structure of a Lie algebra over $\mathbb{R}$ by the
{\it Jacobi bracket} $\{\varphi,\psi\} = \mathbf{E}_{\varphi}(\psi) - \mathbf{E}_{\psi}(\varphi)$.
The {\it algebra of contact symmetries} $\mathrm{Sym}_0 (\EuScript{E})$ is the Lie subalgebra of
$\mathrm{Sym} (\EuScript{E})$ defined as $\mathrm{Sym} (\EuScript{E}) \cap C^\infty(J^1(\pi))$.

Consider $\EuScript{W} = \mathbb{R}^\infty$ with  coordinates $w^s$, $s \in  \mathbb{N}_{0}$. A
{\it differential covering} of $\EuScript{E}$ locally is a trivial bundle
$\tau \colon J^\infty(\pi) \times \EuScript{W} \rightarrow J^\infty(\pi)$
equipped with {\it extended total derivatives}
\[
\widetilde{D}_{x^k} = D_{x^k} + \sum \limits_{ s =0}^\infty
T^s_k(x^i,u^\alpha_I,w^j)\,\frac{\partial }{\partial w^s}
\]
such that $[\widetilde{D}_{x^i}, \widetilde{D}_{x^j}]=0$ for all $i \not = j$ whenever
$(x^i,u^\alpha_I) \in \EuScript{E}$.
Define the par\-ti\-al derivatives of $w^s$ by  $w^s_{x^k} =  \widetilde{D}_{x^k}(w^s)$.  This yields the system
of {\it covering equations}
\begin{equation}
w^s_{x^k} = T^s_k(x^i,u^\alpha_I,w^j)
\label{WE_prolongation_eqns}
\end{equation}
that is compatible whenever $(x^i,u^\alpha_I) \in \EuScript{E}$. Dually, the differential covering is defined by the
{\it Wahlquist--Estabrook forms}
\begin{equation}
d w^s - \sum \limits_{k=1}^{m} T^s_k(x^i,u^\alpha_I,w^j)\,dx^k
\label{WEfs}
\end{equation}
as follows: when $w^s$  and $u^\alpha$ are considered to be functions of $x^1$, ... , $x^n$, forms \eqref{WEfs}
are equal to zero whenever system \eqref{WE_prolongation_eqns} holds.

\subsection{Twisted cohomology of Lie algebras}

For a Lie algebra
 $\mathfrak{g}$ over $\mathbb{R}$, its representation $\rho \colon \mathfrak{g} \rightarrow \mathrm{End}(V)$,
and $k \ge 1$
let $C^k(\mathfrak{g}, V) =\mathrm{Hom}(\Lambda^k(\mathfrak{g}), V)$
be the space of all $k$--linear skew-symmetric mappings from $\mathfrak{g}$ to $V$. Then
the Chevalley--Eilenberg differential
complex
\[
V=C^0(\mathfrak{g}, V) \stackrel{d}{\longrightarrow} C^1(\mathfrak{g}, V)
\stackrel{d}{\longrightarrow} \dots \stackrel{d}{\longrightarrow}
C^k(\mathfrak{g}, V) \stackrel{d}{\longrightarrow} C^{k+1}(\mathfrak{g}, V)
\stackrel{d}{\longrightarrow} \dots
\]
is generated by the differential $d \colon \theta \mapsto d\theta$ such that
\[
d \theta (X_1, ... , X_{k+1}) =
\sum\limits_{q=1}^{k+1}
(-1)^{q+1} \rho (X_q)\,(\theta (X_1, ... ,\hat{X}_q, ... ,  X_{k+1}))
\]
\[
\quad
+\sum\limits_{1\le p < q \le k+1} (-1)^{p+q+1}
\theta ([X_p,X_q],X_1, ... ,\hat{X}_p, ... ,\hat{X}_q, ... ,  X_{k+1}).
\]
The cohomology groups of the complex $(C^{*}(\mathfrak{g}, V), d)$ are referred to as
the {\it cohomology groups of the Lie algebra} $\mathfrak{g}$ {\it with coefficients in the representation}
$\rho$. For the trivial representation $\rho_0 \colon \mathfrak{g} \rightarrow \mathbb{R}$,
$\rho_0 \colon X \mapsto 0$, the cohomology groups are denoted by
$H^{*}(\mathfrak{g})$.

Consider a Lie algebra $\mathfrak{g}$ over $\mathbb{R}$ with non-trivial first cohomology group
$H^1(\mathfrak{g})$ and take a closed 1-form $\alpha$ on $\mathfrak{g}$ such that $[\alpha] \neq 0$.
Then for any $c \in \mathbb{R}$
define new differential
$d_{c \alpha} \colon C^k(\mathfrak{g},\mathbb{R}) \rightarrow C^{k+1}(\mathfrak{g},\mathbb{R})$ by
the formula
\[
d_{c \alpha} \theta = d \theta - c \,\alpha \wedge \theta.
\]
From  $d\alpha = 0$ it follows that
$d_{c \alpha} ^2=0$. The cohomology groups of the complex
\[
C^1(\mathfrak{g}, \mathbb{R})
\stackrel{d_{c \alpha}}{\longrightarrow}
\dots
\stackrel{d_{c \alpha}}{\longrightarrow}
C^k(\mathfrak{g}, \mathbb{R})
\stackrel{d_{c \alpha}}{\longrightarrow}
C^{k+1}(\mathfrak{g}, \mathbb{R})
\stackrel{d_{c \alpha}}{\longrightarrow} \dots
\]
are referred to as the {\it twisted} {\it cohomology groups}  \cite{Novikov2002,Novikov2005} of $\mathfrak{g}$
and denoted by $H^{*}_{c\alpha}(\mathfrak{g})$.


\section{Lie algebra $\qq_2$, its extension, and Lax representations
for equation \eqref{deformed_MASh4}}

As it was shown in \cite{Morozov2019}, the structure equations for the Lie algebra of contact symmetries of
equation \eqref{FKh4} have the form
\begin{equation}
\fl
\left\{
\begin{array}{lcl}
d\alpha &=& 0,
\\
dB &=& \nabla_1(B) \wedge B,
\\
d\Gamma &=& \alpha \wedge \Gamma + \nabla_1(\Gamma) \wedge B +\textfrac{1}{2}\,\nabla_1(B) \wedge \Gamma,
\\
d\Theta &=& \nabla_2(\Theta) \wedge \Theta
+h_0\,\nabla_0(\Theta) \wedge \left( \textfrac{1}{2}\,\nabla_1(B)+h_0\,\nabla_1(\Gamma)-\alpha\right)
\\
&&
+ \nabla_1(\Theta) \wedge (B+h_0\,\Gamma),
\end{array}
\right.
\label{FKh4_se}
\end{equation}
where
\[
B = \beta_0 + h_1\,\beta_1+\textfrac{1}{2}\,h_1^2 \beta_2,
\qquad
\Gamma = \gamma_0 + h_1\,\gamma_1,
\]
\begin{equation}
\Theta =\sum \limits_{k=0}^{2}
\sum \limits_{i=0}^{\infty}\sum \limits_{j=0}^{\infty} \frac{h_0^k h_1^i h_2^j}{i! j!}\, \theta_{k,i,j},
\label{Theta_FKh4}
\end{equation}
and $\alpha$, $\beta_i$, $i \in \{0, 1, 2\}$, $\gamma_l$, $l \in \{0, 1\}$, $\theta_{k,i,j}$,
$k\in \{0, 1, 2\}$, $i, j \in \mathbb{N}_{0}$, are the {\sc mc} forms  of the Lie algebra $\qq_3$,
while $h_0$, $h_1$, $h_2$ are formal parameters such that $dh_i =0$ and $h_0^k = 0$ for $k>2$.

Now we shrink the Lie algebra $\qq_3$ by imposing the condition  $h_0^k =0$ for $k >1$. Thus we have
\begin{equation}
\Theta =\sum \limits_{k=0}^{1}
\sum \limits_{i=0}^{\infty}\sum \limits_{j=0}^{\infty} \frac{h_0^k h_1^i h_2^j}{i! j!}\, \theta_{k,i,j}
\label{Theta_deformed_MASh4}
\end{equation}
instead of \eqref{Theta_FKh4} and
\begin{equation}
\fl
\left\{
\begin{array}{lcl}
d\alpha &=& 0,
\\
dB &=& \nabla_1(B) \wedge B,
\\
d\Gamma &=& \alpha \wedge \Gamma + \nabla_1(\Gamma) \wedge B +\textfrac{1}{2}\,\nabla_1(B) \wedge \Gamma,
\\
d\Theta &=& \nabla_2(\Theta) \wedge \Theta
+h_0\,\nabla_0(\Theta) \wedge \left( \textfrac{1}{2}\,\nabla_1(B)-\alpha\right)
\\
&&
+ \nabla_1(\Theta) \wedge (B+h_0\,\Gamma)
\end{array}
\right.
\label{deformed_MASh4_se}
\end{equation}
instead of \eqref{FKh4_se}.

System \eqref{deformed_MASh4_se} implies that $H^1(\qq_2) = \langle\, \alpha \,\rangle$,
\[
H_{c\,\alpha}^2(\mathfrak{q}_{\diamond}) =
\left\{
\begin{array}{lcl}
\langle\, [\gamma_0 \wedge \gamma_1] \,\rangle, &~~~& c=2,
\\
\{[0]\}, && c \neq 2,
\end{array}
\right.
\]
and $H_{2\,\alpha}^2(\qq_{\diamond}) \subseteq H_{2\,\alpha}^2(\qq_2)$.
Equation
\begin{equation}
d\sigma = 2\,\alpha \wedge \sigma + \gamma_0 \wedge \gamma_1
\label{FKh4_sigma_eq}
\end{equation}
with unknown 1-form $\sigma$ is compatible with the structure equations \eqref{deformed_MASh4_se}. System
\eqref{deformed_MASh4_se},
\eqref{FKh4_sigma_eq} defines the structure equations for the extension
$\widehat{\mathfrak{q}}_2$  of the Lie algebra $\mathfrak{q}_2$.

Frobenius' theorem allows one to integrate equations \eqref{deformed_MASh4_se},
\eqref{FKh4_sigma_eq} step by step. In particular, we have
\[
\alpha=\frac{dq}{q},
\quad
\beta_0 = a_0^2 \,dt,
\quad
\beta_1 = 2\,\frac{da_0}{a_0}+a_1\,dt,
\quad
\beta_2 = \frac{1}{a_0^2}\,\left(da_1 +\textfrac{1}{2}\,a_1^2\,dt\right),
\]
\[
\gamma_0 = a_0\,q\,\,\left(dz+y\,dt\right),
\quad
\gamma_1 = \frac{q}{a_0}\,\left(dy +\textfrac{1}{2}\,a_1\,(dz+y\,dt)\right),
\]
\[
\sigma = q^2\,(dv- y\,dz -\textfrac{1}{2}\,y^2\,dt),
\quad
\theta_{0,0,0} = b_0\,dt+b_1\,dx,
\]
\[
\theta_{1,0,0} = \frac{q\,b_1}{a_0}\,\left(du+b_2\,dt +b_3\,dx+ b_0b_1^{-1}\,dz\right),
\]
where $t$, $x$, $y$, $z$, $u$, $q\neq 0$, $a_0 \neq 0$, $a_1$, $b_0 \neq 0$, $b_1\neq 0$, $b_2$, and $b_3$ are
free parameters (`constants of integration'). We do not need explicit expressions for the other {\sc    mc} forms in
what follows.

Now we {\it impose} the condition for form $\theta_{1,0,0}-\gamma_0$ to be a multiple of
the contact form  $du-u_t\,dt-u_x\,dx-u_y\,dy-u_z\,dz$ on the bundle of jets of sections of the bundle
$\pi \colon \mathbb{R}^4 \times \mathbb{R} \rightarrow  \mathbb{R}^4$,
$\pi \colon (t,x,y,z,u) \mapsto  (t,x,y,z)$, that is, we require
\[
\theta_{1,0,0} -\gamma_0 =\frac{q\,b_1}{a_0}\, (du-u_t\,dt-u_x\,dx-u_y\,dy-u_z\,dz).
\]
To achieve this, we rename the integration parameters as
$b_0 = -u_z\,u_y^{-1}+\textfrac{1}{2}\,a_1$,
$b_1 = u_y^{-1}$, $b_2 = -u_t+\textfrac{1}{2}\,a_1\,y\,u_y$, and
$b_3 = -u_x$.

Then we consider form
\[
\tau = \sigma+c_1\,\beta_0 +c_2\,\gamma_0+c_3\,\gamma_1+c_4\,\theta_{0,0,0}
\]
\[
=q^2 \,\left(dv+ K\,dt+\frac{c_4}{q^2\,u_y}\,dx + \frac{c_3}{a_0 q}\,dy
+\frac{c_3a_1+2\,a_0\,(c-2a_0-q\,y)}{a_0 q}\,dz
\right),
\]
where $K = -2\,c_4a_0\,u_z+2\,c_1a_0^3+2\,c-2a_0^2 y\,q+c-3\,a_1q\,y+c_4a_0a_1-a_0q^2y^2$
and $c_1$, ... , $c_4 \in \mathbb{R}$ are constants. Without loss of generality we put $c_3 = c_4 =- 1$.
Then we rename the integration parameters as
\[
a_0 = -\frac{w}{u_y^{1/2}v_x^{1/2}},
\quad
a_1 = -\frac{2\,w\,(v_z+y-c_2\,w)}{u_yv_x},
\quad
q = \frac{1}{u_y^{1/2}v_x^{1/2}}
\]
and obtain
\[
\tau = u_y\,v_x\,\left(
dv
-(u_z\,v_x+(w+y)\,v_z-(c_2+c_1)\,w^2+y\,w+\textfrac{1}{2}\,y^2)\,dt
\right.
\]
\[
\qquad
\left.
- v_x\,dx-v_z\,dz + u_y\,v_x\,w^{-1}\,dy
\right).
\]
This is the restriction of the multiple of the contact form $dv-v_t\,dt-v_x\,dx-v_y\,dy-v_z\,dz$ from the bundle
of jets of sections of the bundle $\mathbb{R}^4 \times \mathbb{R}^2 \rightarrow \mathbb{R}^4$,
$(t,x,y,z,u,v) \mapsto (t,x,y,z)$, to the submanifold defined by  the over-determined system
\begin{equation}
\left\{
\begin{array}{lcl}
v_t&=& u_z\,v_x+(w+y)\,v_z-(c_2+c_1)\,w^2+y\,w+\textfrac{1}{2}\,y^2,
\\
v_y &=& -u_y\,v_x\,w^{-1}.
\end{array}
\right.
\label{v_covering_1}
\end{equation}
The compatibility condition of this system reads $(v_t)_y = (v_y)_t$.
This equation entails $c_2+c_1=-\textfrac{1}{2}$ and the following system
for $w$:
\begin{equation}
\left\{
\begin{array}{lcl}
w_t&=& u_z\,w_x+(w+y)\,w_z+E\,w\,u_y^{-1},
\\
w_y &=& -u_y\,w_x\,w^{-1}+1,
\end{array}
\right.
\label{w_covering_1}
\end{equation}
where $E = u_{ty} - u_z\,u_{xy}+u_y\,u_{xz}-y\,u_{yz}$.
The compatibility condition $(w_t)_y = (w_y)_t$ of system \eqref{w_covering_1} gives new over-determined
system
\[
\left\{
\begin{array}{lcl}
E_x&=& (u_{xy}+1)\,u_y^{-1}\,E,
\\
E_y&=& u_{yy}\,u_y^{-1}\,E,
\end{array}
\right.
\]
and them condition $(E_x)_y=(E_y)_x$ yields $E=0$, that is, \eqref{deformed_MASh4}.
Substituting this into \eqref{w_covering_1} and renaming $w=p-y$ gives the Lax representation
\begin{equation}
\left\{
\begin{array}{lcl}
p_t&=& \displaystyle{u_z\,p_x+p\,p_z, \phantom{\frac{-u_y\,p_x}{p-y}}}
\\
p_y &=& \displaystyle{\frac{-u_y\,p_x}{p-y}}
\end{array}
\right.
\label{p_covering}
\end{equation}
for equation \eqref{deformed_MASh4}. Notice that $p= \lambda =\mathrm{const}$ is a solution to
\eqref{p_covering}.  Substituting for $w=\lambda -y$, $c_1+c_2 = -\textfrac{1}{2}$ and renaming
$v=r+\textfrac{1}{2}\,\lambda^2\,t$ produces another Lax representation
\begin{equation}
\left\{
\begin{array}{lcl}
r_t&=& \displaystyle{u_z\,r_x+\lambda\,r_z, \phantom{\frac{-u_y\,p_x}{p-y}}}
\\
r_y &=& \displaystyle{\frac{u_y\,r_x}{y-\lambda}}
\end{array}
\right.
\label{r_covering}
\end{equation}
for \eqref{deformed_MASh4}.

\vskip 7 pt
\noindent
{\sc Remark 1}.
Systems \eqref{p_covering} and \eqref{r_covering} are not contact equivalent, instead, they are related in the
following sense, c.f. \cite{PavlovChangChen2009,Krasilshchik2015}: suppose that function $P(t,x,y,z,s)$
defines function $R(t,x,y,z)$ implicitly by equation
\[
P(t,x,y,z,R(t,x,y,z)) \equiv \lambda.
\]
Then $R$ is a solution to \eqref{r_covering} iff $P$ is a solution to \eqref{p_covering}.
\hfill $\diamond$

\vskip 7 pt
\noindent
{\sc Remark 2}.
Parameter $\lambda$ in the Lax representation \eqref{r_covering} is non-removable, that is,
there is no change of variable $r \mapsto \tilde{r}=f(t,x,y,z,r)$ that elinimates $\lambda$.
In accordance with
\cite[\S\S~3.2, 3.4]{KrasilshchikVinogradov1989},
\cite{Krasilshchik2000,IgoninKrasilshchik2000,Marvan2002,IgoninKerstenKrasilshchik2002},
to ensure this claim it is sufficient to notice that
symmetry $\phi = u_y - t\,u_z$ of equation \eqref{deformed_MASh4} has no lift to a symmetry of system
\eqref{r_covering}, while for the associated vector field $V = -\partial_y+t\,\partial_z$ there holds
\[
\mathrm{e}^{\lambda\,V} \left(
dr -u_z\,r_x\,dt-r_x\,dx - \frac{u_y\,r_x}{y}\,dy -r_z\,dz
\right)
\]
\[
\qquad =
dr -(u_z\,r_x+\lambda\,r_z)\,dt-r_x\,dx - \frac{u_y\,r_x}{y-\lambda}\,dy -r_z\,dz.
\]
\hfill $\diamond$


\section{Recursion operator for symmetries}

To construct a recursion operator for equation \eqref{deformed_MASh4} we use the
considerations based on ideas  of \cite{Sergyeyev2015},  cf.
\cite{KrasilshchikKersten1994,KrasilshchikKersten1995,Marvan2002, %
MalykhNutkuSheftel2004,MarvanSergyeyev2012,MorozovSergyeyev2014,KruglikovMorozov2015}
also.  We find shadows for \eqref{deformed_MASh4} in the covering defined by system \eqref{r_covering}.
One of the shadows is $s=r_x^{-1}$. Differentiating \eqref{r_covering} by $x$ and substituting for
$r_x = s^{-1}$ yields another Lax representation
\begin{equation}
\left\{
\begin{array}{lcl}
s_t&=& \displaystyle{u_z\,s_x+\lambda\,s_z-u_{xz}\,s, \phantom{\frac{-u_y\,p_x}{p-y}}}
\\
s_y &=& \displaystyle{\frac{u_y\,s_x-u_{xy}\,s}{y-\lambda}}
\end{array}
\right.
\label{s_covering}
\end{equation}
for \eqref{deformed_MASh4}.
Note that $s$ is a solution to the linearization
\[
\widetilde{D}_t\widetilde{D}_y(\phi) = u_z\,\widetilde{D}_x\widetilde{D}_y(\phi)
+u_{xy}\,\widetilde{D}_z(\phi)-u_y\,\widetilde{D}_x\widetilde{D}_z(\phi)
-u_{xz}\,\widetilde{D}_y(\phi)
\]
\begin{equation}
\qquad
\qquad
\qquad
+y\,\widetilde{D}_y\widetilde{D}_z(\phi)
\label{linearized_deformed_MASh4}
\end{equation}
of \eqref{deformed_MASh4} with the extended total derivatives $\widetilde{D}_t$, ... , $\widetilde{D}_z$
from the covering defined by \eqref{r_covering}.
Now we put
\begin{equation}
s = \sum \limits_{n=-\infty}^{\infty} \lambda^n \,s_n.
\label{s_series}
\end{equation}
Since equation \eqref{linearized_deformed_MASh4} is independent of $\lambda$, each $s_n$ is a solution
to this equation as well. Substitution for \eqref{s_series} into \eqref{s_covering} yields
\[
\left\{
\begin{array}{lcl}
\widetilde{D}_t(s_{n+1})&=& \displaystyle{u_z\,\widetilde{D}_x(s_{n+1})
+\widetilde{D}_z(s_{n})-u_{xz}\,s_{n+1}, \phantom{\frac{-u_y\,p_x}{p-y}}}
\\
y\,\widetilde{D}_y(s_{n+1}) &=& \displaystyle{u_y\,\widetilde{D}_x(s_{n+1})
+\widetilde{D}_y(s_{n})
-u_{xy}\,s_{n+1}
}
\end{array}
\right.
\]
for each $n \in \mathbb{Z}$. Fixing $n$ and renaming $s_{n+1} = \psi$, $s_n = \phi$, we get
the recursion operators $\psi =\EuScript{R}(\phi)$ and $\phi =\EuScript{R}^{-1}(\psi)$ defined by systems
\begin{equation}
\left\{
\begin{array}{lcl}
\widetilde{D}_t(\psi)&=& \displaystyle{u_z\,\widetilde{D}_x(\psi)-u_{xz}\,\psi
+\widetilde{D}_z(\phi), \phantom{\frac{-u_y\,p_x}{p-y}}}
\\
\widetilde{D}_y(\psi) &=& \displaystyle{
\frac{1}{y}\,\left(
u_y\,\widetilde{D}_x(\psi)-u_{xy}\,\psi
+\widetilde{D}_y(\phi)
\right)
}
\end{array}
\right.
\label{rho}
\end{equation}
and
\begin{equation}
\left\{
\begin{array}{lcl}
\widetilde{D}_y(\phi)&=&
\displaystyle{
y\,\widetilde{D}_y(\psi)+u_y\,\widetilde{D}_x(\psi)-u_{xy}\,\psi,
 \phantom{\frac{-u_y\,p_x}{p-y}}
}
\\
\widetilde{D}_z(\phi) &=& \displaystyle{
\widetilde{D}_t(\psi)
+u_z\,\widetilde{D}_x(\psi)
-u_{xy}\,\psi,
}
\end{array}
\right.
\label{inverse_rho}
\end{equation}
respectively.
Direct computations show that systems \eqref{rho} and \eqref{inverse_rho} are compatible iff
$\phi$ and $\psi$ are shadows of symmetries of equation \eqref{deformed_MASh4} in the covering
\eqref{r_covering}.


\section{B\"acklund transformations}

Consider equation \eqref{deformed_MASh4} written in variables
$\tilde{t}$, $\tilde{x}$, $\tilde{y}$, $\tilde{z}$, $\tilde{u}$,
\begin{equation}
\tilde{u}_{\tilde{t}\tilde{y}}  = \tilde{u}_{\tilde{y}} \,\tilde{u}_{\tilde{x}\tilde{z}}
- \tilde{u}_{\tilde{z}}\,\tilde{u}_{\tilde{x}\tilde{y}} +\tilde{y}\,\tilde{u}_{\tilde{y}\tilde{z}},
\label{tilded_deformed_MASh4}
\end{equation}
and make the following point transformation:
\begin{equation}
\tilde{t} = t,
\qquad
\tilde{x} = z,
\qquad
\tilde{y} = q,
\qquad
\tilde{z} = y,
\qquad
\tilde{u} = -x.
\label{first_point_transform}
\end{equation}
The second prolongation of this transformation maps equation \eqref{tilded_deformed_MASh4} to equation
\begin{equation}
q_{yz} = q_{tx} +\frac{q\,q_y-q_t}{q_x}\,q_{xx} +\frac{q_z-q\,q_x}{q_x}\,q_{xy}.
\label{second_modified_FKh4}
\end{equation}
This equation is related to the reduced quasi-classical self-dual Yang--Mills equation
\eqref{FKh4} by a B\"acklund transformation
\begin{equation}
\left\{
\begin{array}{lcl}
q_t&=& \displaystyle{q\,q_y-u_y\,q_x,}
\\
q_z &=& \displaystyle{(q-u_x)\,q_x}.
\end{array}
\right.
\label{second_covering_FKh4}
\end{equation}
Indeed, compatibility condition $(q_t)_z =(q_z)_t$ of system \eqref{second_covering_FKh4} yields
\eqref{FKh4}, while from \eqref{second_covering_FKh4} we have
\begin{equation}
\left\{
\begin{array}{lcl}
u_x&=& q -q_z\,q_x^{-1},
\\
u_y &=& (q_t-q\,q_y)\,q_x^{-1},
\end{array}
\right.
\label{inverse_second_covering_FKh4}
\end{equation}
and then $(u_x)_y = (u_y)_x$ gives equation \eqref{second_modified_FKh4}.
Thus the superposition of transformations  \eqref{first_point_transform} and \eqref{second_covering_FKh4}
provides the B\"acklund transformation between equations \eqref{deformed_MASh4} and \eqref{FKh4}.

As we have shown in \cite{KruglikovMorozov2016}, equation \eqref{FKh4} is related by a B\"acklund
transformation to equation \eqref{MASh4}. Therefore equations \eqref{deformed_MASh4} and \eqref{MASh4}
are related by a B\"acklund transformation as well.
To write this transformation explicitly we substitute \eqref{inverse_second_covering_FKh4} into
\eqref{FKh4_covering_lambda}. The resulting system
\begin{equation}
\left\{
\begin{array}{lcl}
s_t&=& \displaystyle{\lambda\,s_y -\frac{q_t-q\,q_y}{q_x}\,s_x,
\phantom{\frac{\frac{A}{A}}{\frac{A}{A}}}}
\\
s_z &=& \displaystyle{\left(\lambda-\frac{q\,q_x-q_z}{q_x}\right)\,s_x
\phantom{\frac{\frac{A}{A}}{\frac{A}{A}}}}
\end{array}
\right.
\label{BT_1}
\end{equation}
defines a B\"acklund transformation between equation \eqref{second_modified_FKh4} and
\begin{equation}
s_{yz} = s_{tx} +\frac{\lambda\,s_y-s_t}{s_x}\,s_{xx} +\frac{s_z-\lambda\,s_x}{s_x}\,s_{xy}.
\label{first_modified_FKh4}
\end{equation}
Then the point transformation
\begin{equation}
t=\hat{t},
\qquad
x=-\hat{u},
\qquad
y=\hat{z}-\lambda\,\hat{t},
\qquad
z=\hat{x},
\qquad
s = \hat{y}
\label{second_point_transform}
\end{equation}
maps equation \eqref{first_modified_FKh4} to equation \eqref{MASh4} written as
\[
\hat{u}_{\hat{t}\hat{y}}  = \hat{u}_{\hat{y}} \,\hat{u}_{\hat{x}\hat{z}}
- \hat{u}_{\hat{z}}\,\hat{u}_{\hat{x}\hat{y}}.
\]
Thus the superposition of transformations \eqref{first_point_transform}, \eqref{BT_1}, and
\eqref{second_point_transform} maps equation \eqref{deformed_MASh4} to \eqref{MASh4}.

\section*{Acknowledgments}

This work was partially supported by the Faculty of Applied Mathematics of AGH UST statutory tasks within
subsidy of Ministry of Science and Higher Education (Poland).
The author thanks UiT---the Arctic University of Norway and the TFS project Pure Mathematics in Norway for
the financial support of his visit to Troms\o, where a part of the work  was performed.
I am very grateful to professor Boris Kruglikov for his warm hospitality in Troms{\o}.


\end{document}